# Axicons in FSO Systems

Leonardo A. Ambrosio[†], Michel Zamboni-Rached[††], and Hugo E. Hernández-Figueroa[†], *IEEE Senior Member*

[†]Faculdade de Engenharia Elétrica e de Computação (FEEC), Departamento de Microonda e Óptica (DMO), Universidade Estadual de Campinas (UNICAMP), Av. Albert Einstein, 400, 13083-970 – Campinas – SP
[††]Centro de Ciências Naturais e Humanas, Universidade Federal do ABC, Santo André – SP.

*Abstract* — This paper studies the possibility of using axicons in Free Space Optics (FSO) systems. The behavior of the pseudo-Bessel beams generated by "logarithmic" and "linear" axicons, with or without stops, was analyzed through the Huygens-Fresnel integral of diffraction in cylindrical coordinates. We also show that GRIN (Gradient Index) axicons, when well designed, could be used in order to choose the intensity pattern along the propagation axis, which could be a new technique for the alignment equipment.

## I. Introduction

Free-space optical communication systems are still not fully exploited, though this technology arises as an excellent option for several applications, in particular, the so-called last mile. For interconnecting costumers over a last mile link, the advantages of free-space optics (FSO) include ease of implementation, manipulation and mechanical maintenance, low cost compared with fibers, higher bandwidths and no license requirements. Several kilometers can be reached with availabilities up to practically 100 %.

Almost a decade ago, Aruga [1] proposed a method to control the wavefront associated with the transmitted wave using a simple Galilean Telescope. In that work, almost non-diffracting beams (pseudo-bessel beams) were obtained by an appropriate choice of a spherical aberration associated with the diverging lens of the telescope. His subsequent works [2,3] showed that these beams suffered much less atmospheric interferences compared to collimated or focalized beams, and propagation distances up to 50 km were theoretically predicted. In this case, for the same power radiated by the transmitter, these long-range pseudo-bessel beams are advantageous when compared to the conventional Gaussian beams usually adopted in FSO systems.

Although lens with aberrations presents characteristics of extended focus, thus making the reconstruction of the beam along the longitudinal axis of propagation, an improved approach for obtaining non-diffracting patterns would be the use of axicons. Proposed by McLeod in 1954 [4], these optical elements transform the incident wave – usually a plane or Gaussian one - into a Bessel beam, or more generally, a non-diffracting beam with also an extended focus. This happens because the incident beam suffers a change in its wave front, and a non-diffracting pattern is achieved after leaving the conical surface of the axicon such that it propagates without distortion for a maximum distance $z_{max}$, when it decays. In fact, diffraction occurs normally, and the term "non-diffracting" should be taken carefully. The energy propagates from the wave's lateral wings to the optical axis, and it is this lateral energy that reconstructs and maintains the wave's main lobe intact; after the maximum distance, no lateral energy is available for further reconstruction and the wave suddenly collapses.

It can be easily shown that the maximum distance depends on the axicon angle $\gamma$ through the relation $z_{max} = R/\tan\gamma$, $R$ being the axicon radius.

Nevertheless, this is not a general rule, for it is possible to determine the adequate phase function necessary for creating almost any desirable intensity pattern at the optical axis, and such possibilities include the so-called linear and logarithmic axicons. The former can increase the intensity linearly, whereas the latter keeps it uniform along a pre-fixed distance. As an alternative, gradient index axicons [5] – known as GRIN axicons - with an index of refraction that exhibits a specific transverse profile can also accomplish virtually any intensity profile at the longitudinal axis.

This paper presents a theoretical and numerical modeling of these axicons for FSO, based on the Huygens-Fresnel principle for calculating the beam at a certain distance from the transmitter, considered herein as a cylindrical aperture.

## II. Theoretical Analysis

When the distance is much larger than wavelength, we can determine the propagating wave, generated by a finite aperture, at any point in a lossless space by solving the Huygens-Fresnel integral:

$$U(x,y) = \frac{z}{j\lambda} \iint_{S_0} U(\xi,\eta) \frac{e^{jkR}}{R^2} d\xi d\eta \qquad (1)$$

where $U(x,y,z)$ represents a scalar function, which can be any component of the electric or magnetic fields, $U(\xi,\eta)$ is the field distribution at the finite aperture, i.e., at the end of the transmitter, $\lambda$ is the wavelength and $k$ is the wave number associated with the propagating disturbance and $S_0$ is the area of the aperture. Here, $R = \sqrt{z^2 + (x-\xi)^2 + (y-\eta)^2}$ is the distance between a point at the aperture and another one on the lossless medium where $U(x,y,z)$ is to be calculated.

For our purposes, if $R \gg \lambda$ we can expand $R$ in a power series [6]. Let us suppose that the transmitter emits an azimuthally symmetric perturbation with a phase function $\phi(r)$, valid for $\rho_a \leq r \leq \rho$, where $\rho$ is the radius of the aperture (if $\rho_a \neq 0$, then we have an annular aperture), and that we are interested in the intensity of the transmitted wave. As an example, $\phi(r) = -(k/2f)r^2$ for a lens with focal distance $f$, and $\phi(r) = -(1+d^2/\rho^2)^{-1/2} r$ for a plano-convex axicon with focal length $d$. With this in mind, (1) can be used to specify the intensity $I(r',z) = |U(x,y,z)|^2$ at point $(r',z)$, in cylindrical coordinates, as

$$I(r',z) = \left(\frac{k}{z}\right)^2 \left| \int_{\rho_a}^{\rho} \exp\left\{ j\left[\frac{kr^2}{2z} + \phi(r)\right]\right\} J_0\left(\frac{kr'r}{z}\right) r\, dr \right|^2 \quad (2)$$

$J_0$ being the zeroth-order Bessel function, $r'$ the transversal distance to the optical axis.

So, if we want a certain intensity pattern along the optical axis of the FSO system, we should first find the adequate phase function $\phi(r)$ for the problem in question. The actual technology allows the design of almost any required optical element to correctly fit (2), and even though our purpose is focused in this optical system, this theory could be extended to any situation where a certain energy distribution is to be used.

To find the phase function, we shall use the characteristic of energy conversion at the optical axis. Consider an infinitesimal increment $dr$ for the radius of the aperture and, for azimuthally symmetrical problems we may consider, in a first approximation, that the rays leaving this infinitesimal annular ring of width $dr$ cross and longitudinal element $dz$ of the optical axis at a specific point. With no losses, the energy from the ring is transferred to this element $dz$ so that we may write, for the total energy transferred [7],

$$2\pi \int_{\rho_a}^{\rho} P_\sigma(r) r\, dr = \int_{d_1}^{z(r)} P_z(z)\, dz \quad (3)$$

In (3), $P_\sigma(r)$ is the bidimensional power density at the aperture (units of $1/m^2$) and $P_z(r)$ the unidimensional axial power density (units of $1/m$). This is, of course, an idealization, as possible diffraction effects are not taken into account. We could say, then, that $P_z(r)$ is a first-order approximation of the real intensity at the optical axis. The phase function can be found as [8]

$$\varphi(r) = -\int \frac{r}{\sqrt{r^2 + z^2(r)}}\, dr \quad (4)$$

Now, it is a simple task to solve (2), once $P_\sigma(r)$ and $P_z(r)$ are known. Usually, the power density $P_\sigma(r)$ possesses a Gaussian profile, typical of lasers. Assuming that $P_z(r)$ is linear with $z$, we should find a phase function of a typical linear axicon, whereas supposing a constant $P_z(r)$, (4) gives us the phase function of a logarithmic axicon. Although one could be tempting to assume paraxial approximation at this point, so that (4) would be simplified, it is noteworthy that (2) includes longitudinal distances that could be close to the transmitter, so that non-paraxial solutions to the phase functions are to be found.

Suppose that the following pattern for the intensity is desired for a linear axicon with Gaussian profile at the aperture:

$$P_\sigma(r) = P_\sigma\left(1 + b^2 r^2\right)^{-1/2} \text{ and } P_z(z) = cz, \quad (5)$$

where $c$ and $P_\sigma$ are constants, and $b$ is chosen in such a way as to guarantee that $P_\sigma(r)$ is a good approximation of a Gaussian profile (note that this formula is valid if the relation $\Psi = P_\sigma(r=\rho)/P_\sigma(r=0) \geq 0.85$ (this would be equivalent to $d_1/d_2 \geq 0.85$) is satisfied, which allow us to write $b^2 = (1-\Psi^2)/\rho^2 \Psi^2$). For $d_1 \leq z \leq d_2$ we have, after substitution of (5) in (3) and placing $z(r)$ in (4),

$$\phi(r) = -\left\{ \left[ r^2 + A(b^2 r^2 + 1)^{\frac{1}{2}} + \Theta \right]^{\frac{1}{2}} - \frac{Ab}{2} \ln\left| \frac{2}{Ab}\left[ r^2 + A(b^2 r^2 + 1)^{\frac{1}{2}} + \Theta \right]^{\frac{1}{2}} + \frac{2}{Ab^2}(b^2 r^2 + 1)^{\frac{1}{2}} + 1 \right| \right\} + cte \quad (6)$$

In (6), $A = (d_2^2 - d_1^2)/\left[\sqrt{b^2 \rho^2 + 1} - \sqrt{b^2 \rho_a^2 + 1}\right]$ and $\Theta = d_1^2 - A\sqrt{b^2 \rho_a^2 + 1}$. The last term $cte$ can be freely chosen, and we will take it to be such that it makes $\phi(r=0) = 0$, i.e.,

$cte = \left\{ [A+\Theta]^{\frac{1}{2}} - \frac{Ab}{2}\ln\left|\frac{2}{Ab}[A+\Theta]^{\frac{1}{2}} + \frac{2}{Ab^2} + 1\right| \right\}$. Automatically, for an logarithmic axicon, we take $P_z(z) = c$ and the same Gaussian approximation in (5), so that, for $d_1 \leq z \leq d_2$, we have the following phase front :

$$\phi(r) = -\frac{1}{E}\left\{ \left[ Er^2 + C(b^2 r^2 + 1)^{\frac{1}{2}} + \Theta' \right]^{\frac{1}{2}} - \frac{Cb}{2\sqrt{E}} \log\left| \frac{2\sqrt{E}}{Cb}\left[ Er^2 + C(b^2 r^2 + 1)^{\frac{1}{2}} + \Theta' \right]^{\frac{1}{2}} + \frac{2E}{Cb^2}(b^2 r^2 + 1)^{\frac{1}{2}} + 1 \right| \right\} + cte_2 \quad (7)$$

with $A = (d_2 - d_1)/\left[\sqrt{b^2\rho^2 + 1} - \sqrt{b^2\rho_a^2 + 1}\right]$, $E = 1 + A^2 b^2$, $C = 2A\left[d_1 - A\sqrt{b^2\rho^2 + 1}\right]$ and the last term inside the brackets $\Theta' = d_1^2 + 2A^2 + A^2 b^2 \rho^2 - 2Ad_1(b^2\rho^2 + 1)^{1/2}$. As before, $cte_2$ is such that we have $\phi(r=0) = 0$.

## II. SIMULATION RESULTS

The paraxial versions of (6) and (7) were already used in the literature [9,10] for propagation of a few meters, and it was shown that apodization techniques were necessary for eliminating undesired fluctuations of the intensity, whereas central stops (that could be a simple masking opaque disk centered within the axicon, creating an annular-aperture), when well designed, could also serve as a means of smoothing the pattern along the optical axis. For Gaussian intensity profiles at the end of the transmitter, it is even more difficult to achieve a pre-chosen intensity when compared to uniform intensity profiles. The performance of a Galilean telescope, for example, is drastically affected when one passes from uniform to Gaussian profiles.

Lens with spherical aberration generates almost non-diffraction beams due to the phase shift occurred at its edges, which helps reconstructing them. Together with other aberrations, they can be regarded as a special case of (4). Central stops have optimum results when its radius obeys the relation: $r_s = (d_1/d_2)\rho$. This is quite prohibitive in cases when $d_1$ is of the order of $d_2$, since it is equivalent to an unnecessary waste of power. Apodization, however, follows from more complicated equations, and for our purposes, it can be discarded without loss of generality.

Let us consider a common FSO system with $\lambda = 1300$ nm with a transmitter of radius $\rho = 5$ cm without any stops or apodization techniques. Suppose, moreover, that we want a constant intensity pattern between $d_1 = 500$m and $d_2 = 600$m, thus satisfying the condition: $d_1/d_2 \geq 0.85$. The resulting phase function $\phi(r)$ can be appreciated in Fig. 1, while the simulated intensity is shown in Fig. 2. Although the constant profile could not be exactly observed, one could be tempting to compare it with the case of no coupling between an optical element and the transmitter. Fig. 3 shows such results, and a clear advantage, at least when concerning the intensity (we would as well refer to it as the emitted power), is immediately recognized.

Looking at Figs. 2 and 3, we must consider the fact that this almost non-diffracting beam along a pre-established longitudinal distance possesses a peak intensity of almost 27 times that of conventional methods. Obviously, this could be a serious problem when going over exposure limits and safety standards. On the other hand, it is possible to greatly diminish the power emitted by the transmitter.

One of the important features of these beams is their capability of overcoming the effects of atmospheric absorption [3], reconstructing their shapes after being scattered or attenuated by obstacles as aerosols or water drops. The so-called Bessel beams, which are non-diffracting solutions of the scalar wave equation and that can be regarded as special cases of (2) for a specific phase function, for example, are known to have these properties [11].

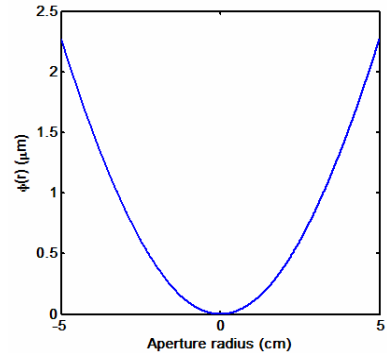

Fig. 1. For a constant intensity profile at 500m < z < 600m, an almost parabolic phase function is needed. Obviously, the more we diminish the difference $d_2 - d_1$, the more $\phi(r)$ resembles a transfer function of a spherical lens.

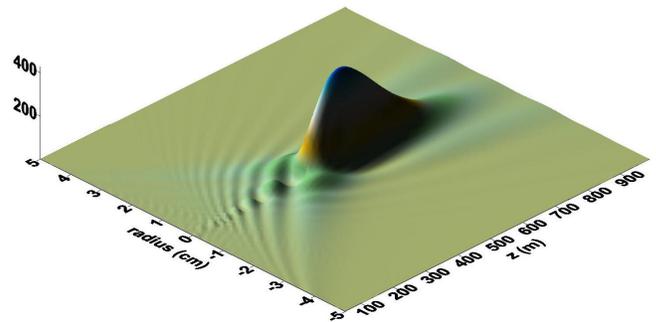

Fig. 2. Intensity profile (arbitrary units), for a logarithmic axicon, supposing a constant pattern at 500m < z < 600m. Due to diffraction, the phase function will not recover the exactly predicted profile.

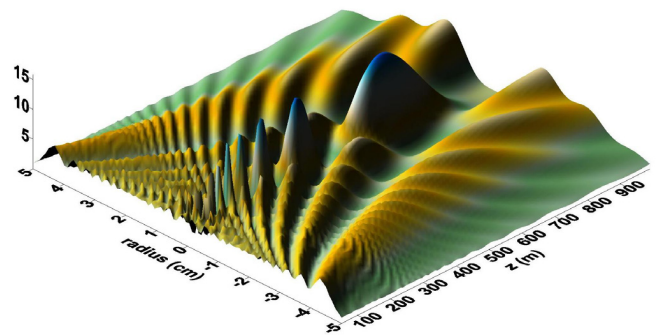

Fig. 3. Intensity profile (arbitrary units), in the absence of an optical element at the transmitter. The peak intensity is almost 27 times smaller than the corresponding one obtained using a logarithmic axicon, see Fig. 2.

We should notice, however, a delicate point about the reliability of using such beams for FSO. As it can be seeing in Fig. 2, these beams are well confined in strict region of space, more precisely, about the longitudinal axis. This is quite disturbing, since, in this type of communication, one usually needs the beam to have some divergence, allowing for possible instabilities that would cause the loss of the line-of-sight. On the other hand, quite collimated beams are used on tracking and adjustment systems so that we could provide this alignment of the system with a simple inclusion of an adequate optical element.

One can find almost the same intensity distribution as in Fig. 2 for a linear axicon, for our approximation cannot predict the exactly solution of (2), i.e., the energy conservation fails for long distances from the aperture. For instance, let us choose a shorter distance, say, 85m < $z$ < 100m, with all the other parameters as before. Taking the constant $c$ of $P_z(r)$ to be $d_1/d_2 = 0.85$, Figs. 4 and 5 show the intensity distribution with and without a stop of $r_s = (d_1/d_2)\rho \approx 4.1667$ cm, respectively, when we couple a linear axicon to the transmitter. In this case, the smoothing process can be clearly observable. An adequate apodization could be further used to enhance the approximation.

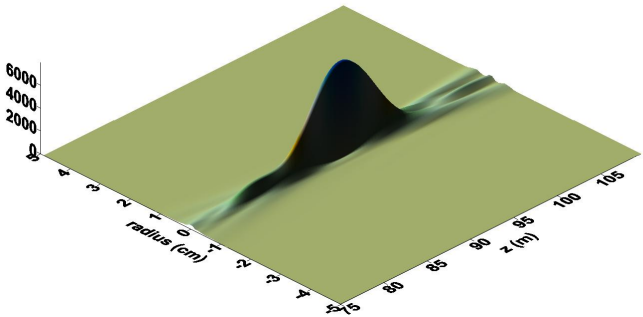

Fig. 4. Intensity distribution (arbitrary units) for a linear axicon having $c = d_1/d_2 = 0.85$ for 85m < $z$ < 100m.

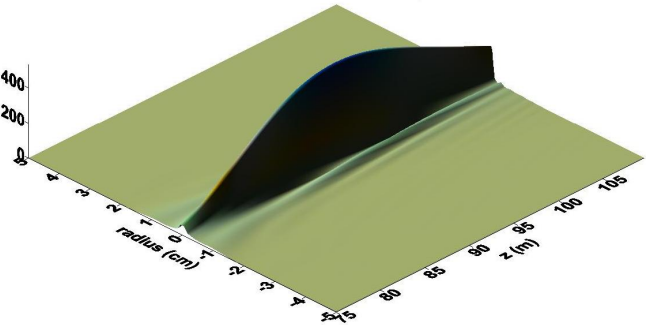

Fig. 5. Even with a stop of $r_s = 4.1667$ cm, the peak intensity is of the order of the intensity with no optical element at all.

An alternative to conventional lens could be used for generating the phase functions just mentioned. These alternatives, also known as GRIN axicons, have the same thickness but different transversal index of refraction, and this is what accounts for the phase difference and, therefore, a specific $\phi(r)$. Once we have established this function, it is straightforward to determine how this GRIN axicon would have to be built. Usually, this can be done by a thin-lens optical path-length argument [12] and, for our purposes, the thin lens index equation can be written as

$$n(r) = n(0) - \frac{\phi(r)}{t} \qquad (8)$$

where $t$ is the thickness of the GRIN axicon and $n(0)$ is the index at its center, known *a priori*.

As a final example, let us make $t$ = 5mm, and $n(0)$ = 1.68. The solution of (8) for the linear axicon of Fig. 4 is the function ploted in Fig. 6. Notice that that the index decay as it goes far from the optical axis. Current technology permits to accurately fabricate this kind of GRIN axicon.

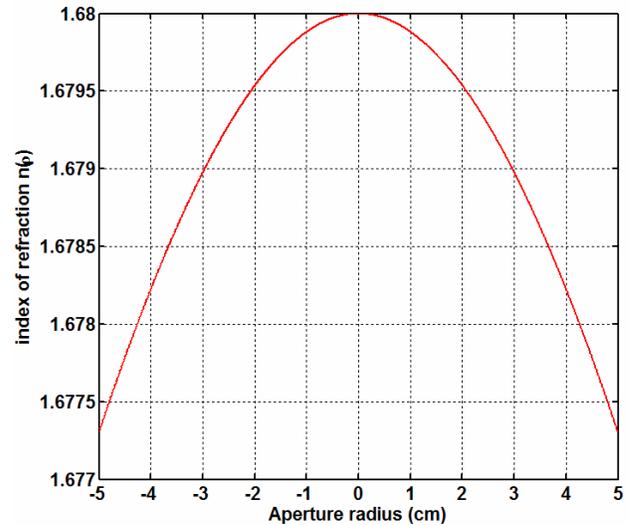

Fig. 6. Index of refraction for a GRIN axicon. It gives the same phase function $\phi(r)$ for the linear axicon of Fig. 4.

III. CONCLUSIONS

A simple theory for designing a phase function that generates almost non-diffracting beams was presented, using the characteristic of energy conservation at the optical axis to predict the intensity profile of these propagating beams. Linear and logarithmic axicons were analyzed using the Huygens-Fresnel principle, and it was shown that, due to diffraction, the approximation of energy conservation cannot predict the real phase function.

The results showed that the peak intensity is increased when compared to other conventional methods, and this implies that one could work with lower emitting power. As for the narrow characteristic of these beams, this comes as another possible alternative for alignment and tracking. Stops and apodization techniques could be used to smooth the intensity and give us a pattern closer to the designed profile.

This work was supported by FAPESP – *Fundação de Amparo à Pesquisa e ao Ensino do Estado de São Paulo*, under contracts 2005/54265-9 (Ph D grant) and 2005/51689-2 (CePOF, Optics and Photonics Research Center).